# Transient drift of *Escherichia coli* under diffusing Step nutrient profile

Sibendu Samanta,[1] Ritwik Layek,[1,*] Shantimoy Kar,[2] Sudipta Mukhopadhyay,[1] and Suman Chakraborty[2, 3, 4*]

[1]*Department of Electronics and Electrical Communication Engineering*
[2]*Advanced Technology Development Centre*
[3]*Department of Mechanical Engineering*
[4]*School of Medical Science and Technology,*
*Indian Institute of Technology Kharagpur, WB-721302, India*

## Abstract

Bacteria such as *Escherichia coli* (*E. coli*) exhibit biased motion if kept in a spatially non-uniform chemical environment. Here, we bring out unique time-dependent characteristics of bacterial chemotaxis, in response to a diffusing spatial step ligand profile. The experimentally obtained temporal characteristics of the drift velocity are compared with the theoretical and Monte-Carlo simulation based estimates, and excellent agreements can be obtained. These results bring in new insights on the time-responsive facets of bacterial drift, bearing far reaching implications in understanding their migratory dynamics in the quest of finding foods by swimming toward the highest concentration of food molecules, or for fleeing from poisons, as well as towards the better understanding of therapeutic response characteristics for certain infectious diseases.

Email: suman@mech.iitkgp.ernet.in and ritwik@ece.iitkgp.ernet.in

Survival of a motile bacterium depends on its response to the environmental fluctuations. Bacterial chemotaxis is one such phenomenon where the single cellular prokaryote, such as *Escherichia coli* (*E. coli*), modulates its movement towards the chemically favourable environment. In a homogeneous medium, the bacteria show Brownian like motion without any specific spatio-temporal bias. However, in presence of spatial chemical concentration gradient, bacteria drift towards the chemical attractants such as glucose, fructose, amino acids and away from the chemical repellents such as ethyl alcohol [1,2]. Moreover, the diffusing spatial step offers a temporal pulse of the nutrient gradient; which in turn generates transient drift that plays an essential role in intermittent bacterial migration. Furthermore, it plays potentially decisive roles in infection and diseases, as attributed to the fact that concentration gradient driven signalling pathways are broadly distributed across a variety of pathogenic bacteria. Estimation of chemotactic features, thus, is essential for the initial stages of infection in different active pathogens.

Various elements of the chemotaxis of *E. coli* [3–7] are depicted in FIG. 1 (see also its caption). Despite recent advancements towards addressing several independent functional modules involved with the same [4,6–8], the time varying response of the drift phenomenon in presence of spatio-temporal fluctuations remains to be properly explained. This deficit emerges from the lack of integration of several complex feedback modules, such as the chemoreceptor adaption, motor adaptation, and feedback between population level drift and the chemoreceptors, exhibiting complex non-linearities and disparate temporal characteristics.

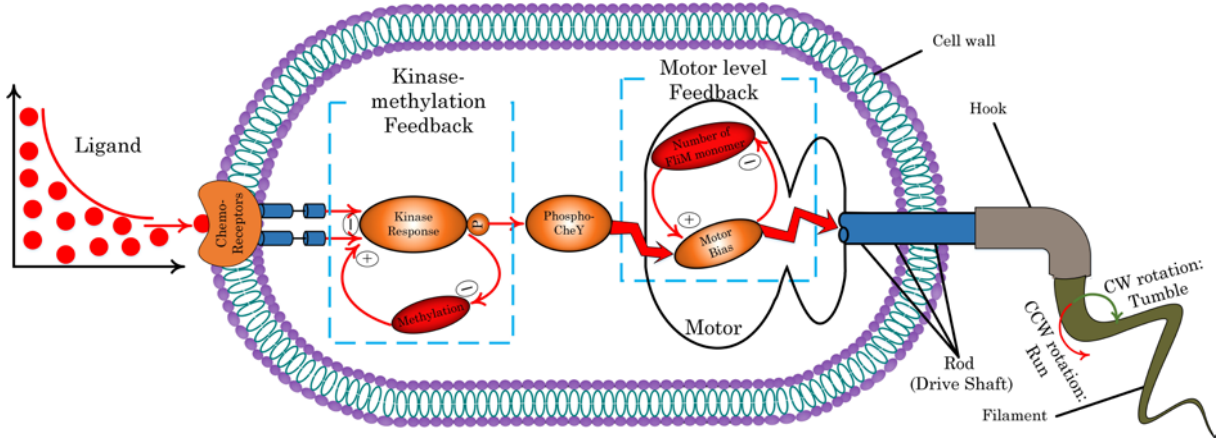

**FIG. 1.** Schematic delineates the chemotaxis pathway of *E. coli*. It broadly includes chemosensory module, signal transduction pathways, and the flagellar motor system. Periplasmic domains of the methyl-accepting chemotaxis proteins (MCP) bind with an attractant/repellent molecule and modulate the auto-phosphorylation activity of the histidine kinase CheA (and CheW) through their cytoplasmic domains. The receptor's adaptation with temporal ligand fluctuation is carried out by a methylation based negative integral feedback loop involving the proteins, CheR and CheB. The phosphorylated CheA-CheW molecules perform signal transduction by phosphorylating the kinase CheY. Thereafter, phospho-CheY molecules relocate themselves form the sensor region to the motor region, bind to the protein FliM, and modulate the CW bias of flagellar motor. The motors also can adapt to a certain extent to the changes in the intracellular phospho-CheY level by varying the composition (number of FliM monomers) of their C-rings. During the default counterclockwise (CCW) rotation of the flagellar motor, flagella bundle together to generate decisive propulsion force to drive the bacterial *run*. Whenever some of the motors switch their rotation direction to clockwise (CW), the flagella come out of the bundle and thereby bacterium *tumbles* to get a new direction for its subsequent *run*.

Here, we unveil previously unaddressed transient characteristics of bacterial chemotaxis by providing a unified perspective, focussing on the interrelation between various time scales [8–10] governing the overall process. We discuss the integration for different network topologies, such as the series, the feed-forward parallel, and the feedback topologies to capture the underlying mechanism of spatio-temporal dynamics, consistent with experimentally obtained results. We further validate our theory with comprehensive experiments.

### *Modelling*

FIG. 2 depicts the essential elements associated with chemotactic signalling. Three feedback loops namely: (i) receptor feedback ($F_1$) [11,12], (ii) motor feedback ($F_2$) [6,7], and (iii) drift to receptor feedback ($F_3$) [4] are considered in this integrated model. In the chemoreceptor module, kinase activity $a(t)$ is enhanced by methylation $m(t)$, while the rate of methylation is suppressed by $a(t)$. The pathways for this negative feedback loop [$m(t) \to a(t)$ ⊣ $m(t)$; where → and ⊣ denote the activation and inhibition of a process, respectively] facilitate receptor-level adaptation. Therefore, the major contribution of this feedback loop lies in desensitization of the steady state kinase activity in presence of temporal ligand fluctuations. During motor level adaptation [6,7], the bacterial flagellar motor can adapt to the partial alterations of intracellular phospho-CheY concentration by altering its own composition. Due to the bacterial drift towards high concentration of attractant molecules, the ligands get attached to the receptors, which in turn decreases the concentration of the corresponding ligand in that spatial location. Thus, over a population of bacteria, this phenomenon provides a feedback from the drift ($v_d$) to the receptors' methylation transport rate [4,12].

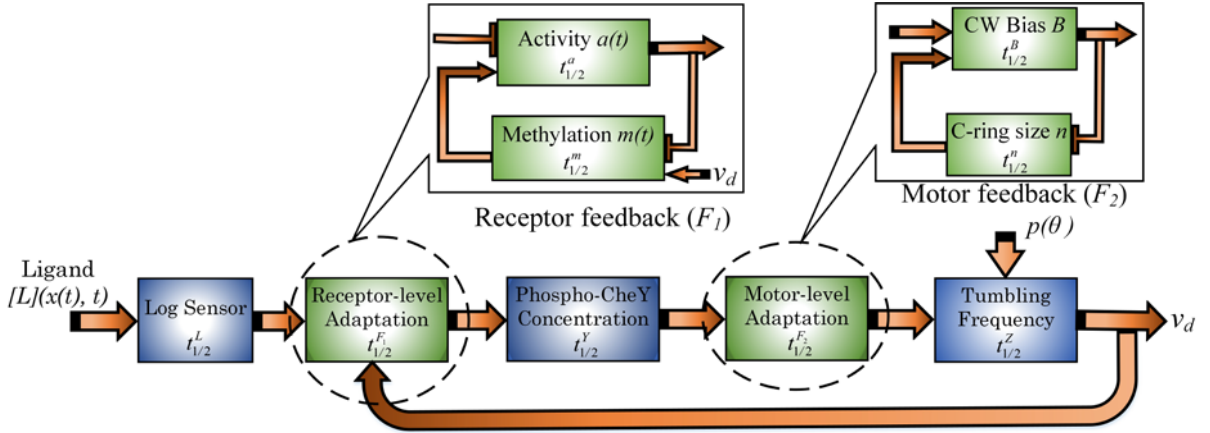

**FIG. 2.** Block diagram illustrates the independent modules and corresponding interconnections with defined time responses. Receptor feedback: [$F_1$: $m(t) \rightarrow$ kinase activity ($a(t)$) $\dashv m(t)$], motor feedback: [$F_2$: $n \rightarrow$ motor CW bias ($B$) $\dashv n$], and drift to receptor feedback: $\left(F_3 : v_d \rightarrow \frac{dm(t)}{dt}\right)$. $t_{1/2}^{v_d}$ is half-time (i.e. time taken by specific variable to reach its half-maximal value) for $F_3$ path (i.e. path from $v_d$ to receptor-level adaptation module). Here, $t_{1/2}^{a}$ and $t_{1/2}^{m}$ are half-time of kinase activity $a(t)$ and methylation $m(t)$ path for $F_1$, respectively. Similarly, $t_{1/2}^{B}$ and $t_{1/2}^{n}$ are half-time of motor-bias $B$ and motor feedback ($n$) path for $F_2$. $t_{1/2}^{F_1}$ and $t_{1/2}^{F_2}$ are the half time of negative feedback topologies $F_1$ and $F_2$. $t_{1/2}^{Y}$ and $t_{1/2}^{Z}$ are half-time for phosphorylating the CheY and *tumbling* time, respectively. $t_{1/2}^{L}$ is the binding half-time of ligand to receptor. (Details in Section 1 of *Supplementary Material* [19]).

Motion of an *E. coli* in a two-dimensional space is schematically shown in FIG. 3. Ligand concentration at any specific location ($x(t)$, $y(t)$) is given by *[L]($x(t)$, $t$)*, while the spatial ligand variation is assumed to be in one dimension (*x*-direction). The bacterium starts its *(k+1)th run* at time $t_k$ from the location ($x_k$, $y_k$). In Monod-Wyman-Changeux (MWC) model [13], *E. coli's* chemotactic sensor module can mainly be divided into two parts: logarithmic-sensors that sense the external nutrient concentrations of exponential variability, and intracellular methylation based receptor adaptation. The output of this sensor receptor complex is modelled by a single variable $a(t) = 1/(1 + e^{N(f_{[L]} + f_m)})$ [3,14] that depends upon the cooperative behaviour of $f_m$ and $f_{[L]}$ and it is normalized between 0 and 1. Here, $N$ is the average number of ligand-binding units in each receptor and $f_m = \alpha(m_o - m(t))$ is the methyl-dependent average free energy per binding site characterized by the initial value of methylation $m_o$ (methylation level when no ligand is bound to it), scaling parameter $\alpha$, and methylation level $m(t)$ at time $t$. $f_{[L]} = ln\left[1 + \frac{[L](x(t),t)}{K_i}\right] - ln\left[1 + \frac{[L](x(t),t)}{K_a}\right]$ is ligand-dependent average free energy (per binding site) which is characterized by the dissociation constants $K_a$ and $K_i$ for active and inactive state of the receptors respectively. In the effective range of ligand concentration (i.e. $K_i < [L](x(t),t) < K_a$), ligand-dependent average free energy becomes $f_{[L]} \approx ln\left[\frac{[L](x(t),t)}{K_i}\right]$. The dynamics of methylation rate ($dm(t)/dt$) is controlled by two feedback loops ($F_1$ and $F_3$) involving the receptor kinase activity $a(t)$ and drift velocity $v_d$ as: $dm(t)/dt = \frac{1}{\alpha}\left[v_d G + \frac{1}{Na(t)[1-a(t)]}\frac{da(t)}{dt}\right]$, where $G = \frac{1}{[L](x(t),t)}\frac{\partial [L](x(t),t)}{\partial x}$ is the spatial derivative of ligand concentration per unit ligand concentration.

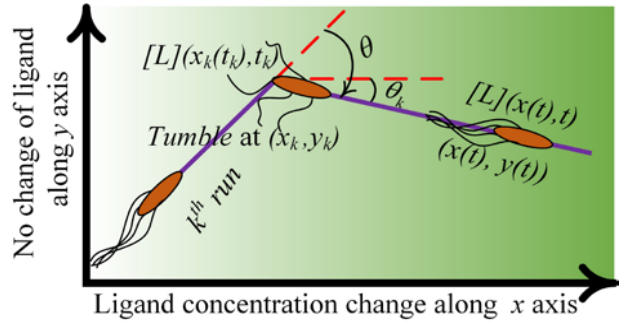

**FIG. 3.** Schematic representation of the bacterial *run-tumble* motion in two dimensional space where ligand concentration changes only in *x*-direction. The net *tumbling* angle $\theta_k$ with the positive *x*-axis is essentially the cumulative sum of all *k tumbling* angles ($\theta$) and the angle of initial *run* ($\theta_o$) with the positive *x*-axis.

Active receptor (phospho-CheA-CheW complex) starts downstream signal transduction by phosphorylating CheY. Phospho-CheY relocates itself towards the motor module and binds to FliM [15,16]. Whenever the accumulation of phospho-CheY ($[Y]$) crosses a threshold limit, rotor switches to its direction of CW with bias value $B = (1 + \lambda\, exp(n\,\chi\,))^{-1}$; where $\lambda$ is the equilibrium constant for transitions between CW and CCW states (in absence of the ligand), $\chi = ln\left[\frac{1 + [Y]/K_{ccw}}{1 + [Y]/K_{cw}}\right]$ is a non-uniform scaling of $[Y]$ with two dissociation constants $K_{cw}$ and $K_{ccw}(>K_{cw})$. $K_{cw}$ and $K_{ccw}$ are related to binding of $[Y]$ to FliM during the CW and CCW phases respectively. Also, $n$ is the number of FliM monomers in the C-ring of flagellar motor. CW rotation of the motor untangles the flagellar bundle resulting in a *tumble*, which randomly chooses a new direction for subsequent *run*. Following *tumble*, the motor resets itself to its default CCW state and continues its *run*. According to MWC model [6,7,13], flagellar motor's C-ring encircles a ring of FliM monomers ($n$) which varies between $n_{cw}$ and $n_{ccw}$, depending on the CW bias ($B$) of the motor. Flagellar motor can partially adapt to changes in $[Y]$ by modulating the number of FliM monomers ($n$) in C-ring. The existence of auto-regulatory negative feedback mechanism with the FliM monomers ($n$) is given by $\frac{dn}{dt} = k_o\left[\frac{(1-B)}{1 + \gamma\Delta n/(n_{ccw}-n)} - \frac{B}{1 + \gamma\Delta n/(n-n_{cw})}\right]$; where $\Delta n = n_{ccw} - n_{cw}$. Here, $k_o$ represents adaptation speed of the motor and $\gamma$ ($<1$) is a fraction of the boundary region where the sizes of C-ring during the CW and CCW phases match.

Motor rotational bias modulates the bacterial effective *tumbling* frequency $\left[ Z = f\left(B, p(\theta)\right) = b\,ln\left[\dfrac{eB}{1-B}\right] + Z_\theta \right]$ [10,15,16]; $Z_\theta$ is rotational diffusion coefficient parameterized on the *tumbling* angle $\theta$, which follows the probability density function: $p(\theta) = \left|0.25(1+cos\,\theta)sin\,\theta\right|$; $\forall\,\theta \in (-\pi, \pi]$ [17,18]. Here, $b$ and $e$ are the scaling and curve fitting parameters [10,15,16] of $Z$, respectively. Whenever there is a spatial gradient of ligand concentration, bias of CW rotation of the motor is reduced and as a result, *tumbling* becomes less frequent, which promotes its drift towards higher ligand concentration. The spatial derivative of the average *run-time* ($1/Z$) is used to compute the chemotactic drift velocity $v_d$ for a population of bacteria swimming against the influence of attractant gradient $\left(\dfrac{\partial[L](x(t),t)}{\partial x}\right)$: $v_d(t) = v_o^2\,\dfrac{d}{dx}\left(\dfrac{1}{Z}\right)$; where $v_o$ is the intrinsic velocity during *run*. In absence of any spatial ligand concentration gradient, *run-time* ($1/Z$) becomes constant and thus chemotactic drift ceases to exist.

Now, it is important to probe the transient characteristics of the interacting dynamical system modules. To understand the response time of the integrated system, firstly several network topologies have been analysed. It can be clearly seen that such transients depend extensively on the chosen topology. As shown in *Supplementary Material* (Section 1) [19], every module has its characteristics half-time. The tentative range of the half-time ($\tau$) of the combined module can be estimated from the half-times of all the constituent modules while each of them is considered to be independently stable.

Half-time ($t_{1/2}^m$) for the methylation path of $F_1$ is derived under the assumption of linearized transfer characteristics of the methylation rate: $t_{1/2}^m = \displaystyle\int_{m_{min}}^{m_{half}} \dfrac{e^{\alpha Nm} + e^{N(u+\alpha m_o)}}{K_R e^{N(u+\alpha m_o)} - K_B e^{\alpha Nm}}\,dm = \left[\dfrac{m}{K_R} - \dfrac{(K_B + K_R)}{\alpha N K_B K_R}ln\left(\left|K_B e^{\alpha Nm} - K_R e^{N(u+\alpha m_o)}\right|\right)\right]_{m_{min}}^{m_{half}}$; where $m_{half} = (m_{min} + m_{max})/2$ is methylation level at half-time. $m_{min}$ and $m_{max}$ are the minimum and maximum value of methylation level. Here, $K_R$ and $K_B$ are the methylation and demethylation rate of the inactive and active receptor complex. Similarly, half-time ($t_{1/2}^n$) for motor's feedback path of $F_2$ is inferred as: $t_{1/2}^n = \displaystyle\int_{n_{min}}^{n_{half}} (1 + \lambda\,e^{n\chi})\,k_o^{-1}\left[\dfrac{\lambda(n_{ccw}-n)e^{n\chi}}{n_{ccw}-n+\gamma\Delta n} - \dfrac{n-n_{cw}}{n-n_{cw}+\gamma\Delta n}\right]^{-1}\,dn$; where, $n_{half} = \left(\dfrac{n_{min}+n_{max}}{2}\right)$ is the C-ring size during half-time. $n_{min}$ and $n_{max}$ are the minimum and maximum value of C-ring size that are $n_{cw}$ and $n_{ccw}$, respectively.

As response time of a negative feedback system is in between its forward and feedback path's half-times, the half-times of $F_1$ ($t_{1/2}^{F_1}$) and $F_2$ ($t_{1/2}^{F_2}$) will be:

$min(t_{1/2}^a, t_{1/2}^m) < t_{1/2}^{F_1} < max(t_{1/2}^a, t_{1/2}^m)$ and $min(t_{1/2}^B, t_{1/2}^n) < t_{1/2}^{F_2} < max(t_{1/2}^B, t_{1/2}^n)$, respectively. The equivalent half-time of a series topology is greater than the half-times of all the constitute modules. So, half-time $t_{1/2}^{series}$ of series topology i.e. the combination of two feedback topologies ($F_1$ and $F_2$), phosphorylation CheY, and *tumbling* frequency block, is $t_{1/2}^{series} > max[t_{1/2}^{F_1}, t_{1/2}^{F_2}, t_{1/2}^{Y}, t_{1/2}^{Z}]$. $F_3$ feedback system's characteristic time $t_{1/2}^{F_3}$ is analysed using negative feedback topology technique: $min(t_{1/2}^{series}, t_{1/2}^{v_d}) < t_{1/2}^{F_3} < max(t_{1/2}^{series}, t_{1/2}^{v_d})$, where, $t_{1/2}^{v_d}$ is half-time of $v_d \to dm(t)/dt$ path. The half-time $\tau$ of the integrated system is deduced by following similar series topology: $\tau > max[t_{1/2}^{F_3}, t_{1/2}^{L}]$. Typical values of individual module' half-time are displayed in Table S1, *Supplementary Material* [19] and we obtain $\tau$ to be in the range of 10-30s; which is consistent with our simulation and experiments. (For more details about different topologies' half-times please see Section 1, *Supplementary Material* [19].)

Next, we explore Monte Carlo simulation to validate the theoretical findings from the proposed model. The diffusing spatial step ligand concentration profile is assumed and 1000 bacteria are considered to begin their *run* from 1000 uniformly chosen initial position in a two-dimensional space. The *run-times* are derived based on our model. However, at every *tumble* location, *tumbling* angle ($\theta$) is assumed to be independent of *[L](x(t), t)*. The angles are drawn from the prescribed probability density function, $p(\theta) = |0.25(1 + cos\theta) sin\theta|$; $\forall \theta \in (-\pi, \pi]$ using Monte Carlo inverse technique [20]. For each bacterium, $10^5$ run-tumble sequences are simulated to obtain the trajectories. The drift velocities, measured for every trajectory, are averaged to obtain the comparison with theoretical estimate.

### Results

Experimental settings used are schematically represented in FIG. 4A and the microfluidic setup made of PDMS is shown in *Supplementary Material* (Section 3) [19]. A 'Y'-shaped microfluidic platform [$l = 30$ mm, $h = 25$ $\mu m$, $w = 800$ $\mu m$; *E. coli*, *DH5α* ($10^6$cfu/ml) suspended in PBS and dextrose (3mM) were transported at $1\mu l$/min through 'B' and 'D' reservoirs respectively] was used to validate the developed theoretical understandings. Dimension of the channels is larger than the bacterial dimension by at least one order of magnitude, so that bacterium can have free access within the micro-conduit.

To investigate the bacterial motility characteristics, we wait for 10s prior to the image acquisition (in stopped flow condition) and thus ensure minimal inertial effect. Distribution of dextrose concentration profile is shown in FIG. 4B. Bright field phase contrast microscopy (OLYMPUS IX 71, 40X objective) is used to capture the bacterial transport at the interfacial region (i.e. along the width of the channel from centrally dotted interface shown in FIG. 4A). *Tumbling* position ($x_k$, $y_k$) is ascertained from the bacterium's trajectory at instant $t_k$ using image processing. The angle between the $k^{th}$ *run* and positive $x$-axis at the time interval ($t_{k-1}$, $t_k$) is given by $\theta_k = \lim_{\Delta x_k \to 0} \tan^{-1} \frac{\Delta y_k}{\Delta x_k} \approx \tan^{-1} \frac{(y_{k+1} - y_k)}{(x_{k+1} - x_k)}$. Variation of *tumbling* angles at every *tumbling* position is depicted in FIG. 5, which suggests the continuous adaptation of motor's feedback mechanism in its translational trajectories.

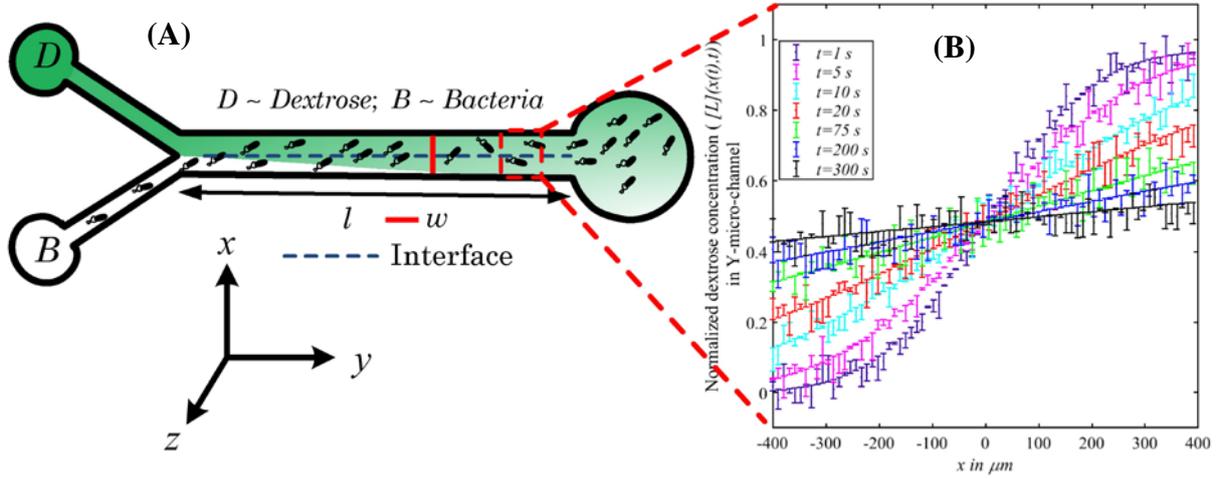

**FIG. 4(A).** Schematic of the 'Y-microchannel' used for experiments: dextrose solution flowing from 'D' reservoir getting diluted by PBS buffer flowing from the 'B' reservoir; and **(B)** Dynamic ligand (dextrose) concentration profile in the 'Y-microchannel': error bars are defined by $\mu \pm 2\sigma$, where $\mu$ and $\sigma$ are mean and standard deviation of the measured values. (Details in Section 3, FIG. S10–S11, *Supplementary Material* [19]).

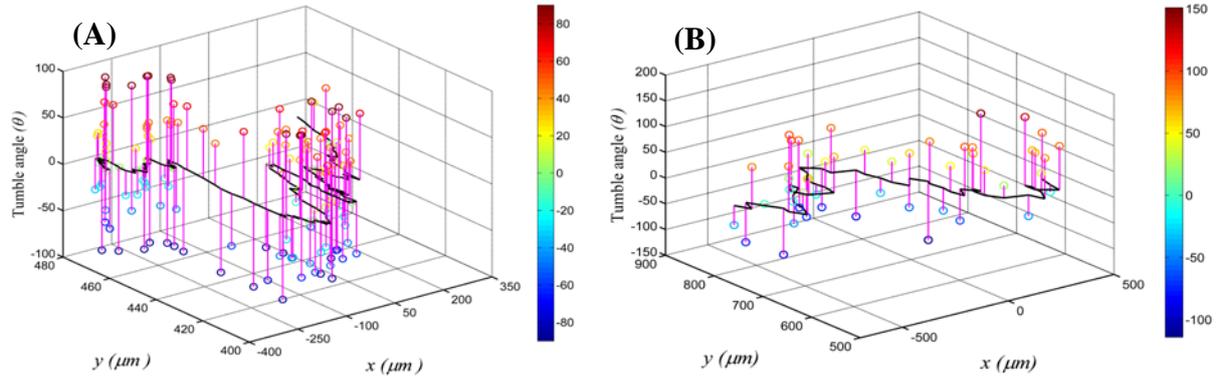

**FIG. 5.** Randomness of *tumbling* angle dictates the *run-tumble* motion; as it is evident from the bacterial trajectories (A) experimental and (B) simulation monitored till 150s. As a representative case, trajectory for only one bacterium is presented.

From FIG. 6A, it is clearly evident that the observed drift can also be precisely predicted from simulation studies. For the specific spatial ligand concentration gradient (in FIG. 4B), there is a steep rise of the drift and a maximum $v_d \sim 4.9 \pm 0.8$ μm/s around 60 seconds; and thereafter it slows down to approximate zero in 300 seconds. In a spatio-temporal ligand field, every *run* corresponds to two components of motion, firstly the random diffusion and secondly the biased *run-time* modulation resulting in drift. If the nutrient concentration profile is maintained in only one direction (e.g. *x* axis), drift is expected to occur in that specified direction only. Hence, the diffusion which is modelled using the *tumbling* angle distribution; can be averaged out when the cumulative angular shift (movement in 2-D space) crosses *2πk* angle (where *k* is an integer).

To rationalize the outcome of the proposed model, a comparative study is presented in FIG. 6B with the reported models (details is discussed in <span style="color:red">Section 2, *Supplementary Material* [19]</span>). From FIG. 6B, it can be easily realized that the individual modules ($F_1$, $F_2$, and $F_3$) or combination of any two individual ones are not substantial for precise estimation of transient drift.

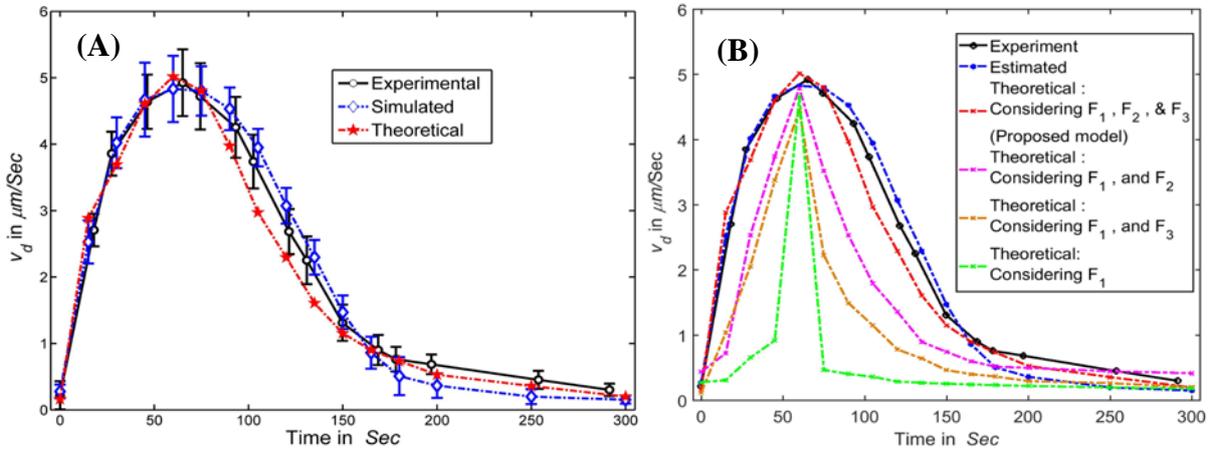

**FIG. 6:** **(A)** Drift velocities measured at different time instances. Experimental and simulation results are in agreement with theoretical evaluations. Error bars are estimated by $\mu \pm 2\sigma$ ; $\mu$ and $\sigma$ are mean and standard deviation of the measurements (22 experimental and 100 simulation observations were considered at every instance of time); and **(B)** Drift velocity $v_d$ at different instant of time in 'Y-microchannel'. Proposed theoretical and estimated results from the simulated bacterial trajectory are used for comparing with estimated result from the experimented bacterial trajectory in 'Y-micro-channel'. We derive theoretical drift velocity based on $F_1$: receptor-level adaptation $\left(m(t) \to a(t) \dashv m(t)\right)$, $F_2$: motor level adaptation $\left(n \to B \dashv n\right)$, and $F_3$: dependence of methylation level dynamic on $v_d$ $\left(v_d \to \dfrac{dm(t)}{dt}\right)$. Outcome of the proposed model (i.e. integrating $F_1$, $F_2$, and $F_3$) shows closest match to the experimental and simulation findings.

***Conclusions and Outlook***

A comprehensive model of bacterial chemotaxis is developed in this manuscript. The three feedback loops are justified from the experimental validations of the transient drift. The quantitative analysis of the response time of the integrated model conforms to the experimental findings as well. Our new results towards resolving the transients of bacterial chemotactic drift are likely to play a fundamental role towards understanding a wide gamut of processes, ranging from disease pathogenesis, biofilm formation, bioremediation, to carbon cycling in the ocean.

# *Supplementary Material*: Transient drift of *Escherichia coli* under diffusing Step nutrient profile

Sibendu Samanta, Ritwik Layek, Shantimoy Kar, Sudipta Mukhopadhyay, and Suman Chakraborty

## Section 1: Transient analysis of different types of network topology

The overall half-time (i.e. the time taken by a variable to reach of its half-maximal value) of a topology is always different from its constituent modules' half-time. Here, we have analyzed the resultant half-time for different types of network topologies (e.g. series, parallel, and negative feedback topology) that have helped to derive overall half-time of the proposed chemotaxis network.

### 1.1 Series topology

A series topology is shown in FIG. S1.

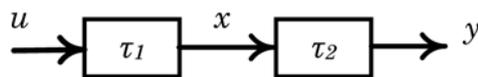

**FIG. S1:** Series topology

Assuming linearity, the dynamic equations describing the series topology can be written as

$$\dot{x} = -\alpha_1 x + \beta_1 u$$
$$\dot{y} = -\alpha_2 y + \beta_2 x; \quad where \ \alpha_1, \beta_1, \alpha_2, \beta_2 > 0 \tag{S1}$$

The half-times of each block are $\tau_1 = \dfrac{ln\,2}{\alpha_1}$ and $\tau_2 = \dfrac{ln\,2}{\alpha_2}$ respectively. The resultant half-time ($\tau$) of the series topology can be derived from Eq. S2. Here, $\tau/\tau_2$ is plotted for different value of $\tau_1/\tau_2$ in FIG. S2.

$$\tau_1 2^{-\tau/\tau_1} - \tau_2 2^{-\tau/\tau_2} = \tfrac{1}{2}(\tau_1 - \tau_2) \tag{S2}$$

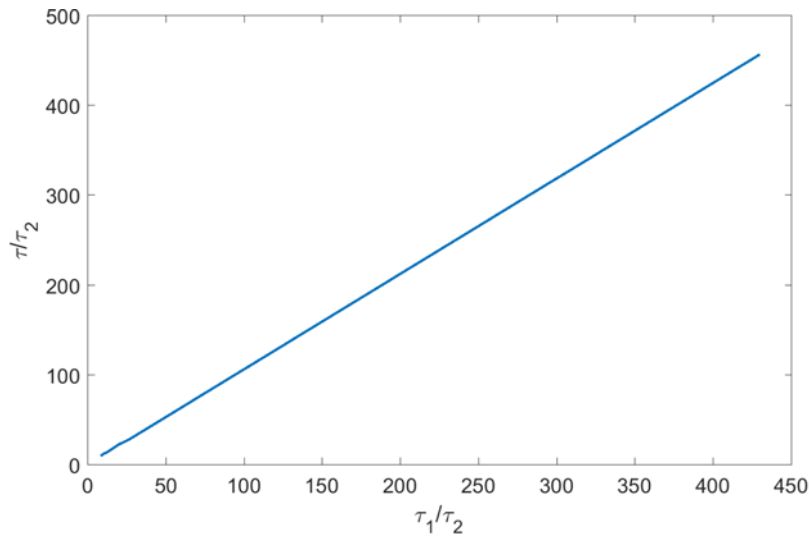

**FIG. S2:** Resultant half-time $\tau$ is plotted for $\tau_1 \in [\,13.6, 693.1\,]\,s$ and $\tau_2 = 1.6$. Figure shows that the resultant characteristic time $\tau > \{\,\tau_1, \tau_2\,\}$.

## 1.2 Parallel topology

A parallel topology is shown in FIG. S3.

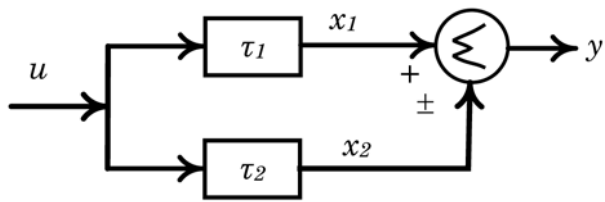

**FIG. S3:** Parallel model.

Similarly, the dynamic equations that describe the parallel topology can be written as

$$\begin{aligned}
\dot{x}_1 &= -\alpha_1 x_1 + \beta_1 u \\
\dot{x}_2 &= -\alpha_2 x_2 + \beta_2 u \\
y &= x_1 \pm x_2 \,; \qquad where\ \alpha_1,\ \alpha_2,\ \beta_1,\ \beta_2 > 0
\end{aligned} \tag{S3}$$

The half-times of each block are $\tau_1 = \dfrac{ln\,2}{\alpha_1}$ and $\tau_2 = \dfrac{ln\,2}{\alpha_2}$ respectively. The resultant half-time ($\tau$) of the parallel topology that can be derived from Eq. S4 for addition and Eq. S5 for subtraction model. For additive parallel topology, $min(\,\tau_1, \tau_2\,) < \tau < max(\,\tau_1, \tau_2\,)$. For subtractive parallel topology, $\tau < \{\,\tau_1, \tau_2\,\}$ when $\tau_1/\tau_2 < 1$ and $\tau > \{\,\tau_1, \tau_2\,\}$ when $\tau_1/\tau_2 > 1$.

$$\beta_1 \tau_1 2^{-\tau/\tau_1} + \beta_2 \tau_2 2^{-\tau/\tau_2} = \tfrac{1}{2}(\,\beta_1 \tau_1 + \beta_2 \tau_2\,) \tag{S4}$$

$$\beta_1 \tau_1 2^{-\tau/\tau_1} - \beta_2 \tau_2 2^{-\tau/\tau_2} = \tfrac{1}{2}(\,\beta_1 \tau_1 - \beta_2 \tau_2\,) \tag{S5}$$

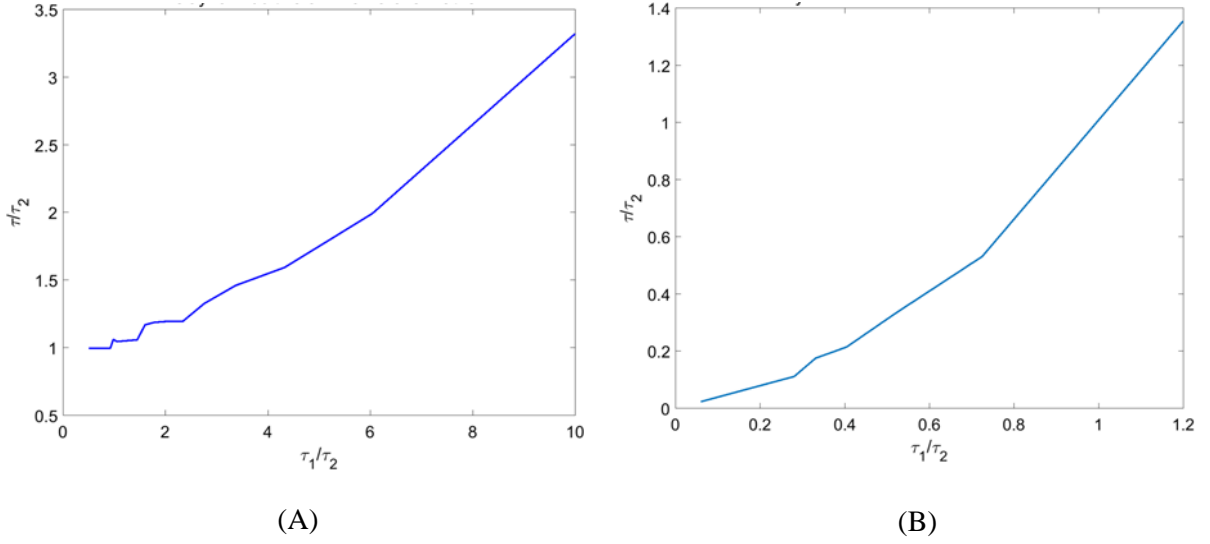

(A)                                                          (B)

**FIG S4:** Half-time is plotted for (A) additive and (B) subtractive parallel topology.

## 1.3    Negative feedback topology:

A negative feedback topology is given in FIG. S5.

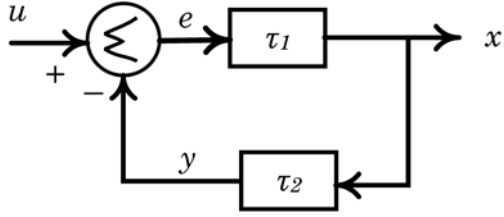

**FIG. S5:** Negative feedback topology.

In the same way, the dynamic equations describing the negative feedback topology can be defined in Eq. S6 and derived using Cayley-Hamilton theorem.

$$
\begin{aligned}
\dot{x} &= -\alpha_1 x + \beta_1 e \\
\dot{y} &= -\alpha_2 y + \beta_2 x \\
e &= u - y; \qquad where \ \alpha_1, \ \alpha_2, \ \beta_1, \ \beta_2 > 0
\end{aligned}
\tag{S6}
$$

$\tau_1 = \dfrac{ln\,2}{\alpha_1}$ and $\tau_2 = \dfrac{ln\,2}{\alpha_2}$ are the half-times for forward block and feedback block, respectively.

The resultant half-time $(\tau)$ of the negative feedback topology that can be derived using Cayley-Hamilton theorem:

$$
(\tfrac{\lambda_1 + \alpha_1}{\lambda_2})e^{\lambda_2 \tau} - (\tfrac{\lambda_2 + \alpha_1}{\lambda_1})e^{\lambda_1 \tau} = \tfrac{1}{2}\left(\tfrac{\lambda_1 + \alpha_1}{\lambda_2} - \tfrac{\lambda_2 + \alpha_1}{\lambda_1}\right)
\tag{S7}
$$

where, $\lambda_{1,2} = \dfrac{-(\alpha_1 + \alpha_2) \pm \sqrt{\alpha_1^2 + \alpha_2^2 - 2\alpha_1\alpha_2 - 4\beta_1\beta_2}}{2}$. The derived $\tau$ using Eq. S7 is plotted in FIG. S6 and $\tau$ will be in between of $\tau_1$ and $\tau_2$.

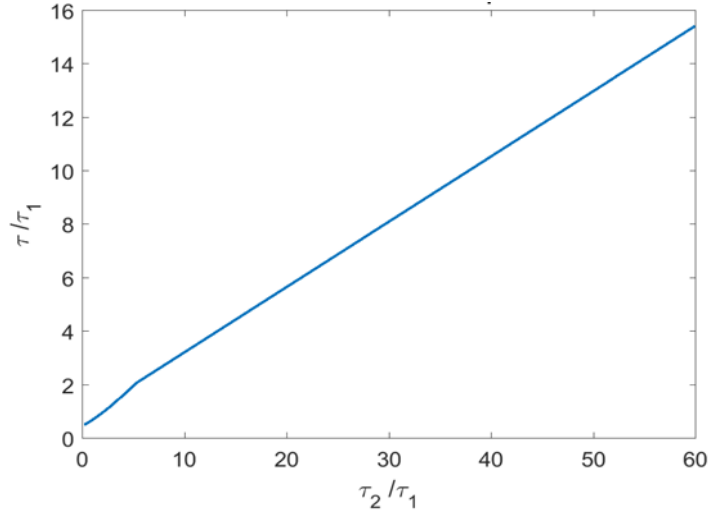

**FIG. S6:** The resultant half-time for negative feedback model. We considered $\tau_1 = 1.1552s$, $\tau_2 \in [\,0.24, 70\,]\,s$ and we get $\tau \in [\,0.59, 17.83\,]\,s$. We get $min(\tau_1, \tau_2) < \tau < max(\tau_1, \tau_2)$.

## 1.4 Our proposed network:

Two negative feedback topologies (methylation level adaptation ($F_1$) and motor level adaptation ($F_2$)) are in series form in our intended network (FIG. S7). $t_{1/2}^{F_1}$ and $t_{1/2}^{F_2}$ are the half-time of these negative feedback topologies (i.e. $F_1$ and $F_2$). $t_{1/2}^{F_1}$ and $t_{1/2}^{F_2}$ are derived using Eq. S7, and we got $min(t_{1/2}^a, t_{1/2}^m) < t_{1/2}^{F_1} < max(t_{1/2}^a, t_{1/2}^m)$ and $min(t_{1/2}^B, t_{1/2}^n) < t_{1/2}^{F_2} < max(t_{1/2}^B, t_{1/2}^n)$. Here, $t_{1/2}^a$ and $t_{1/2}^m$ are half-time of kinase activity and methylation path for $F_1$ – feedback system, respectively. Similarly, $t_{1/2}^B$ and $t_{1/2}^n$ are half-time of motor-bias and motor feedback path for $F_2$ – feedback system. The half-time of the series topology is $t_{1/2}^{series}$ that is derived using its constituent modules' half-time (i.e. $t_{1/2}^{F_1}$, $t_{1/2}^Y$, $t_{1/2}^{F_2}$, and $t_{1/2}^Z$) and we got $t_{1/2}^{series} > \{ t_{1/2}^{F_1}, t_{1/2}^Y, t_{1/2}^{F_2}, t_{1/2}^Z \}$ using Eq. S2. Here, $t_{1/2}^Y$ and $t_{1/2}^Z$ are half-time for phosphorylating CheY and *tumbling* time, respectively. Half-time for $F_3$ – feedback system, $t_{1/2}^{F_3}$ is deduced using its constituent modules' half-time (i.e. $t_{1/2}^{series}$ and $t_{1/2}^{v_d}$) and we got $min(t_{1/2}^{series}, t_{1/2}^{v_d}) < t_{1/2}^{F_3} < max(t_{1/2}^{series}, t_{1/2}^{v_d})$ after solving the Eq. S7. Here, $t_{1/2}^{v_d}$ is half-time of $v_d \rightarrow \frac{dm(t)}{dt}$ path. The resultant half-time $\tau$ of this network is derived using its constituent modules' half-time (i.e. $t_{1/2}^L$, and $t_{1/2}^{F_3}$) and we get $\tau > \{ t_{1/2}^{F_3}, t_{1/2}^L \}$ using series topology technique's half-time calculation (i.e. Eq. S2). Half-time of ligand *[L](x(t),t)* binding to chemoreceptor is defined by $t_{1/2}^L$. Several half-time values are displayed in Table S1 and we obtain $\tau$ to be in the range of 10s to 30s. It is consistent with our findings of the transient analysis.

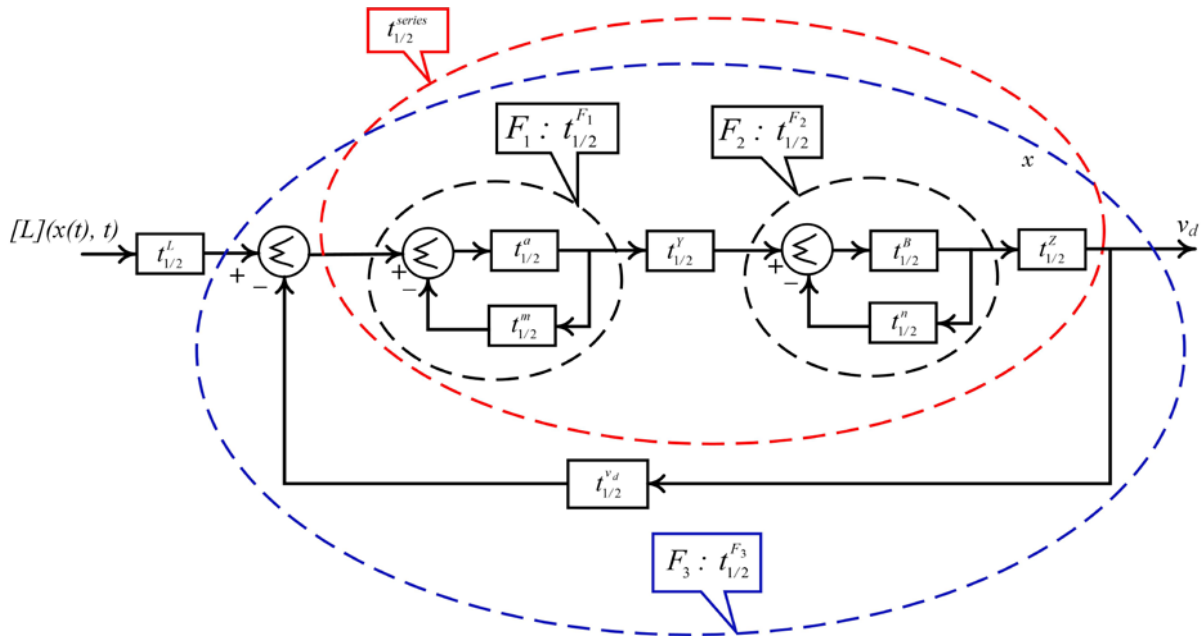

**FIG S7:** Schematic illustration of the proposed half-time model for *E. coli*'s chemotaxis module. We obtain: $min(t_{1/2}^a, t_{1/2}^m) < t_{1/2}^{F_1} < max(t_{1/2}^a, t_{1/2}^m)$, $min(t_{1/2}^B, t_{1/2}^n) < t_{1/2}^{F_2} < max(t_{1/2}^B, t_{1/2}^n)$, $t_{1/2}^{series} > \{t_{1/2}^{F_1}, t_{1/2}^y, t_{1/2}^{F_2}, t_{1/2}^z\}$, $min(t_{1/2}^{series}, t_{1/2}^{v_d}) < t_{1/2}^{F_3} < max(t_{1/2}^{series}, t_{1/2}^{v_d})$, and $\tau > \{t_{1/2}^{F_3}, t_{1/2}^L\}$ using Eq. S2 and S7. $\tau$ is the resultant half-time of our intended network. The typical half-time values of all blocks are displayed in Table S1 and we obtain $\tau$ to be in the range of 10s to 30s. It is consistent with our findings of the transient analysis.

**Table S1:** Value of the half-time of each path

| Half-time | Value | Half-time | Value |
|---|---|---|---|
| $t_{1/2}^L$ | ~1 ms [14,15] | $t_{1/2}^n$ | ~60 s to 100 s [10,13] |
| $t_{1/2}^a$ | ~1 s [14,15] | $t_{1/2}^{F_2}$ | ~20 s to 60 s [10,13] |
| $t_{1/2}^m$ | ~7 to 10 s [14,15] | $t_{1/2}^{v_d}$ | ~60 s to 120 s [3] |
| $t_{1/2}^{F_1}$ | ~2 s to 4 s [14,15] | $t_{1/2}^{series}$ | ~1s to 80s |
| $t_{1/2}^y$ | ~3 to 5 s [16] | $t_{1/2}^{F_3}$ | ~ 5 s to 20 s |
| $t_{1/2}^B$ | ~100 ms to 120 ms  [17] | $\tau$ | ~10 s to 30 s [10,13] |
| $t_{1/2}^Z$ | ~0.1 s | | |

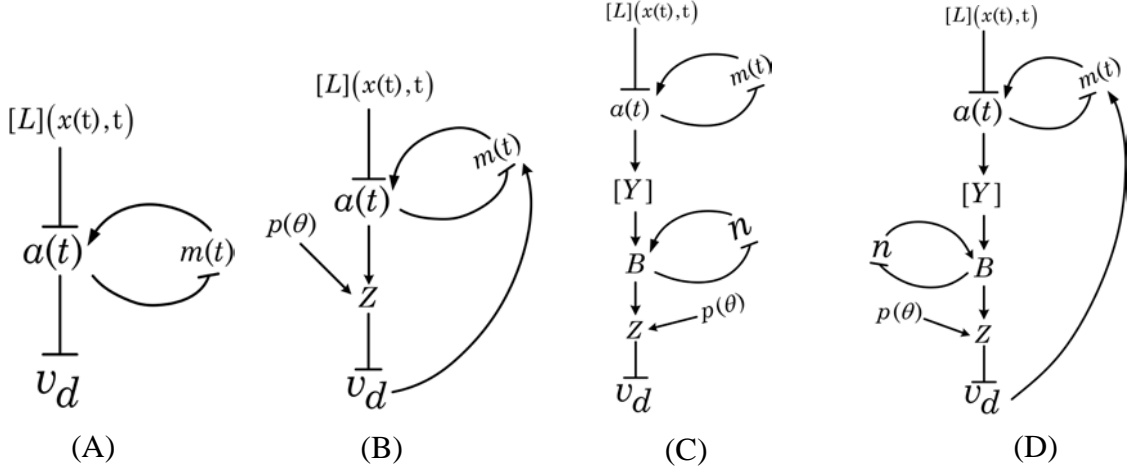

**FIG. S8:** Different types of chemotaxis pathway. $F_1$: receptor level adaptation, $F_2$: motor level adaptation, and $F_3$: dependency of methylation rate on bacterial motility. (A) Chemotaxis pathway with $F_1$, (B) Chemotaxis pathway with $F_1$ and $F_3$, (C) Chemotaxis pathway with $F_1$ and $F_2$, and (D) Chemotaxis pathway with integrate $F_1$, $F_2$, and $F_3$ (Proposed model). Symbols → and ⊣ represent the activation and inhibition of a process, respectively.

## Section 2:  Different types of chemotaxis model for drift velocity

### 2.1  Drift velocity using receptor level adaptation ($F_1$)

Jiang *et al.* [1] (model in FIG. S8(A)) have considered the receptor level adaptation $F_1$: $m(t) \rightarrow activity\ a(t) \dashv m(t)$ [in Eq. (S8)-(S9)] to define the drift velocity $v_d$ that is derived in Eq. (S10). The drift velocity for ligand concentration *[L](x(t),t)* (FIG. S11) is shown in FIG. 6(B) of the main manuscript. After sensing the ligand *[L](x(t),t)* by chemoreceptor, the conformation of the histidine kinase (CheA-CheW) is modulated as Eq. (S8).

$$a(t) = \frac{1}{1+e^{N\left[ln\left(\frac{1+\frac{[L](x(t),t)}{K_i}}{1+\frac{[L](x(t),t)}{K_a}}\right)+\alpha(m_o-m(t))\right]}} \tag{S8}$$

$$F\left(a(t)\right) = \frac{dm(t)}{dt} = \left(1-a(t)\right)K_R - a(t)K_B . \tag{S9}$$

The initial condition and the scaling of the methylation level of the receptor-complex are parameterized by $m_o$ and $\alpha$, respectively. $N$ is the number of ligand-binding units in every receptor. $K_a$ and $K_i$ are the dissociation constants for the active and inactive receptors, respectively. The methylation rate $\left(dm(t)/dt\right)$ of Eq. (S9) is a monotonically decreasing function over the range of $a(t)$. $F(0) = K_R$ is maximum methylation rate by the methyltransferase CheR and $F(1) = -K_B$ is the maximum demethylation rate by the methylesterase CheB. There are several choices possible for the functional $F(a(t))$, the linear one is the most intuitive and meaningful. However, it is possible to experimentally estimate $F(a(t))$ as well. In this model, Eq. (S9) is used as the choice of the methylation rate [2].

$$v_d = \frac{C_m G}{1 + G/G_{cs}} \, ; \tag{S10}$$

where $G = \frac{\partial}{\partial x} ln([L](x(t),t))$, $C_m$ is motility constant of the bacteria, and $G_{cs}$ is a critical gradient beyond which $v_d$ saturates. For $G < G_{cs}$, the drift velocity is linearly dependent on the spatial gradient of the logarithmic ligand concentration: $v_d \approx C_m G$. For $G > G_{cs}$, the drift velocity will be constant ($v_d \approx C_m G_{cs}$).

## 2.2    Drift Velocity: considered receptor adaptation ($F_1$) and dependency of methylation rate on bacterial motility ($F_3$)

In this model (in FIG. S8(B)), receptor level adaptation $\left[ F_1 \colon m(t) \to \text{activity } a(t) \dashv m(t) \right]$ (in Eq. (S11)) and dependency of methylation rate on bacteria motility $\left[ F_3 \colon v_d \to \frac{dm(t)}{dt} \right]$ [in Eq. (S12)] [3] are considered to design the chemotaxis model for the drift velocity.

$$a(t) = \frac{1}{1 + e^{N\left[ ln\left( \frac{1 + \frac{[L](x(t),t)}{K_i}}{1 + \frac{[L](x(t),t)}{K_a}} \right) + \alpha\left(m_o - m(t)\right) \right]}} \simeq \frac{1}{1 + e^{N ln\left[ \frac{[L](x(t),t)}{K_i} \right] + \alpha N(m_o - m(t))}}$$

$$= \frac{1}{1 + e^{N\left[ u(x(t),t) + \alpha\left(m_o - m(t)\right) \right]}} \, ; \; for \; K_i << [L]\left( x(t), t \right) << K_a \tag{S11}$$

$$\frac{dm(t)}{dt} = \frac{v_d}{\alpha} \cdot \frac{\partial u(x(t),t)}{\partial x} + \frac{1}{\alpha Na(t)(1 - a(t))} \cdot \frac{da(t)}{dt} \simeq v_d \frac{G}{\alpha} + \frac{1}{\alpha Na(t)(1 - a(t))} \cdot \frac{da(t)}{dt} \, ;$$

$$where \; G = \frac{\partial}{\partial x}\left( ln\left[ L \right]\left( x(t), t \right) \right); \; u(x(t),t) = ln\left[ \frac{[L]\left( x(t),t \right)}{K_i} \right]. \tag{S12}$$

The kinases activity $a(t)$ changes the concentration of phospho-CheY *[Y]*. The phospho-CheY molecules transport through the cytoplasm and bind with the rotor of the flagellar motor. Whenever the accumulation of phospho-CheY (*[Y]*) crosses a particular threshold, the rotor switches its direction to CW and the flagellar bundle unfolds. Rotational diffusion decides new direction during *tumble* and then the rotor is reset to its default CCW direction to start its next *run*. The effective *tumbling* frequency $Z$ [3–5] can be defined in term of the receptor activity $a(t)$ and rotational diffusion coefficient of bacteria $Z_\theta$ which is a function of *tumbling* angle $\theta$. The $\theta$ follows probability density function $p(\theta)$ in Eq. (S19).

$$Z = Z_o \left( a/a_o \right)^\varsigma + Z_\theta \, ; \tag{S13}$$

where $\varsigma(\approx 10)$ is the Hill coefficient, $Z_o$ is the average *tumbling* frequency at $a = a_o$. The chemotaxis drift velocity is proportional to the spatial derivative of $Z^{-1}$ and it can be written as:

$$v_d(t) = v_o^2 \, \frac{d}{dx}\left(\frac{1}{Z}\right). \tag{S14}$$

The derived drift velocity using this model for particular ligand concentration *[L](x(t),t)* (FIG. S11) is shown in FIG. 6(B) of the main manuscript.

## 2.3 Drift Velocity: considered receptor adaptation (*F₁*) and motor level adaptation (*F₂*)

In this chemotaxis model [FIG. S8(C)], receptor level adaptation $\left[F_1 : m(t) \to \text{activity } a(t) \dashv m(t)\right]$ in Eq. (S8)-(S9) and motor level adaptation $\left[F_2 : n \to \text{motor CW bias } B \dashv n\right]$ [6] are regarded to design $v_d$. The kinases activity ($a(t)$) has changed the concentration of phospho-CheY (*[Y]*) (Eq. (S15)) that followed Hill function [7–9] with hill coefficient ($w_1$) and half-maximal effect ($K_{1/2}$).

$$[Y] = f(a(t)) = \frac{\left(a(t)/K_{1/2}\right)^{w_1}}{1 + \left(a(t)/K_{1/2}\right)^{w_1}} \tag{S15}$$

The phospho-CheY molecules transport through the cytoplasm and bind with FliM and FliN proteins of C-ring. Recently, experiments [6,10] show that the bacterial flagellar motor can adapt to changes in the intracellular level of phospho-CheY by changing its composition. A model is constructed for motor-level adaptation which is modulated the number of FliM monomers in the C-ring. The motor level feedback is governed by FliM exchange between CCW and CW states. Specifically, the rates at which the FliM monomer are switched 'ON'/'OFF' depend upon the 'CW' and 'CCW' states of the motor. The constructed model shows good agreement with the observed motor-level adaptation. In this model, a Monod-Wyman-Changeux (MWC) [11] model can be used to describe the motor's highly cooperative switching behavior (CW bias *B*) that is derived in Eq. (S16).

$$B = \frac{\left(1 + [Y]/K_{cw}\right)^n}{\left(1 + [Y]/K_{cw}\right)^n + \lambda\left(1 + [Y]/K_{ccw}\right)^n} = \frac{1}{1 + \lambda\dfrac{\left(1 + [Y]/K_{ccw}\right)^n}{\left(1 + [Y]/K_{cw}\right)^n}} = \frac{1}{1 + \lambda e^{n \ln\left[\frac{1 + [Y]/K_{ccw}}{1 + [Y]/K_{cw}}\right]}} = \frac{1}{1 + \lambda e^{n\chi}} ; \tag{S16}$$

where $\chi = ln\left[\dfrac{1 + [Y]/K_{ccw}}{1 + [Y]/K_{cw}}\right]$. $K_{cw}$ and $K_{ccw}$ ($>K_{cw}$) are dissociation constants for binding of phospho-CheY to FliM for CW and CCW state, respectively. $\lambda$ is the equilibrium constant for transitions between the CW and CCW states. $n \in [n_{cw}, n_{ccw}]$ is the size of the FliM ring and the motor-level adaptation dynamics are characterized by the function $H$ that can thus be expressed explicitly as:

$$H = \frac{dn}{dt} = k_o\left[\frac{(1 - B)}{1 + \gamma \Delta n / (n_{ccw} - n)} - \frac{B}{1 + \gamma \Delta n / (n - n_{cw})}\right]; \tag{S17}$$

where, $\Delta n = (n_{ccw} - n_{cw})$. $n_{cw}$ and $n_{ccw}$ ($> n_{cw}$) define the size of C-ring for CW and CCW state, respectively. $k_o$ determines the adaptation speed of the motor and $\gamma$ ($<1$) is the fraction of the boundary region where ring size of C-ring for CW and CCW meet. Total effective *tumbling* frequency is indicated by $Z$ and inverse of $Z$ (i.e., $Z^{-1}$) is the nothing but average modulated *run-time*. The definition of $Z$ [8,9,12,13] is given by Eq. (S18). For increasing attractant or decreasing repellent gradient, the *tumbling* frequency decreases from its original value and bacteria stretch their movement toward a favorable direction.

$$Z = f\left(B, p(\theta)\right) = b \, ln\left[\frac{eB}{1-B}\right] + Z_\theta \; ; \tag{S18}$$

where $b$ and $e$ define curve the scaling and fitting parameters of $Z$ respectively. $Z_\theta$ is the rotational diffusion co-efficient that has been affected by the directional fluctuation of bacteria (i.e., distribution of *tumbling* angle, $p(\theta)$ that is defined in Eq. (S19)).

$$p(\theta) = \begin{cases} -0.25(1+cos\,\theta)sin\,\theta & \forall\,\theta \in (-\pi,\,0\,] \\ 0.25(1+cos\,\theta)sin\,\theta & \forall\,\theta \in (\,0,\,\pi\,] \end{cases} \tag{S19}$$

Hence, for the given ligand field *[L](x(t),t)*, the chemotactic drift velocity ($v_d$) is proportional to the spatial derivation of average *run-time* ($Z^{-1}$) and it can be written as Eq. (S20). For particular ligand concentration *[L](x(t),t)* (FIG. S11), the derived drift velocity for this model is shown in FIG. 6(B) of the main manuscript.

$$v_d(t) = v_o^2 \frac{d}{dx}\left(\frac{1}{Z}\right). \tag{S20}$$

## 2.4 Comprehensive Model for Drift Velocity: considered receptor adaptation ($F_1$), motor level adaptation ($F_2$), and dependency of methylation rate on bacterial motility ($F_3$) (*Proposed model*)

For modeling drift velocity $v_d$, all three feedbacks are considered: (1) receptor level adaptation $\left(F_1: m(t) \rightarrow activity\,a(t) \dashv m(t)\right)$ that is defined in Eq. (S11), (2) motor level adaptation $\left(F_2: n \rightarrow motor\,CW\,bias\,B \dashv n\right)$ that has been explained in Eq. (S15) – (S17), and (3) dependence of methylation rate on bacterial motility $\left(F_3: v_d \rightarrow \frac{dm(t)}{dt}\right)$ that is derived in Eq. (S12). The pathway of this model is shown in FIG. S8(D) and corresponding proposed control block diagram of the chemotaxis network is demonstrated in FIG. S9. The drift velocity is derived using Eq. (S18) – (S20). For particular ligand concentration *[L](x(t),t)* (FIG. S11), the drift velocity for this model is shown in FIG. 6(B) of the main manuscript. FIG. 6(B) of the main manuscript shows that none of the individual modules (i.e., $F_1$, $F_2$, and $F_3$,) and the combination of any two modules are capable enough to estimate transient drift under a diffusing step profile of attractant. The integration of the three feedback loops creates a comprehensive model to study bacterial chemotaxis and the instantaneous drift phenomenon.

**Table S2:** Value of the parameters considered in the computational model

| Parameters | Value | Parameters | Value |
|---|---|---|---|
| $v_o$ | 20 $\mu$m/s | $\lambda$ | $10^4$ [3,6] |
| $N$ | 6 [18] | $K_{cw}$ | 1 [3,6] |
| $K_i$ | 18.2 $\mu$M [18] | $K_{ccw}$ | 3.47 [3,6] |
| $K_a$ | 3 mM [18] | $n_{cw}$ | 30 [3,6] |
| $\alpha$ | 1.7 [18] | $n_{ccw}$ | 38 [3,6] |
| $m_o$ | 1 [18] | $k_o$ | 1 [3,6] |
| $a_o$ | 0.33 [18] | $\gamma$ | 0.05 [3,6] |
| $\tau_{run}$ | 1s | $b$ | 0.185 ($\in$ [0.08, 0.2]) [9] |
| $\tau_{tumble}$ | 0.1s | $e$ | 152 ($\in$ [100, 200]) [9] |
| $K_{1/2}$ | 1/3 [9] | $Z_\theta$ | 0.14 ($\in$ [0.1, 0.3]) [3,5] |
| $w_l$ | 6.5 [9] | $D$ | $5.7 \times 10^2$ $\mu$m$^2$/s @ 25$^o$C |

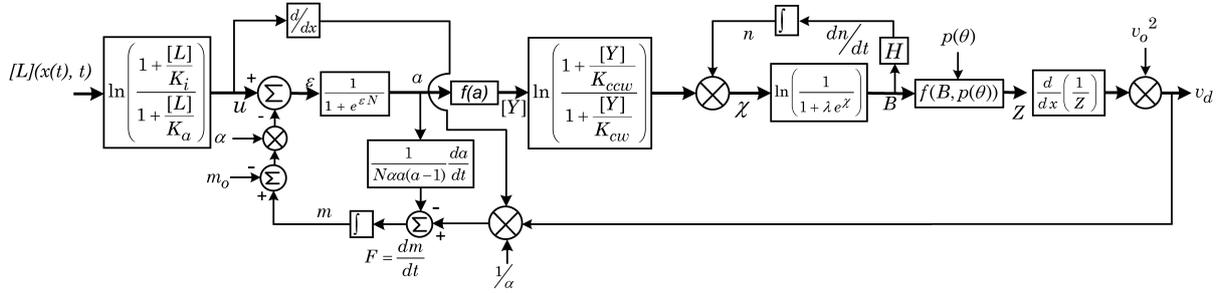

**FIG. S9:** Proposed control block diagram of *E. coli*'s chemotaxis network. Where,

$$\varepsilon = u + \alpha(m_o - m), \quad f(a) = \frac{(a/K_{1/2})^{w_l}}{1 + (a/K_{1/2})^{w_l}}, \quad H = f(B) = \frac{dn}{dt} = k_o \left[ \frac{1-B}{1 + \gamma \, \Delta n/(n_{ccw} - n)} - \frac{B}{1 + \gamma \, \Delta n /(n - n_{cw})} \right],$$

$\Delta n = (n_{ccw} - n_{cw})$, and $Z = f(B, p(\theta)) = b \, \ln \left[ \dfrac{eB}{1-B} \right] + Z_\theta$.

## Section 3    Experimental details

The microfluidic setup made of PDMS is shown in FIG. S10. A 'Y'-shaped microfluidic platform ($l$ =30 mm, $h$ =25 $\mu$m, $w$ =800 $\mu$m) was used to validate the developed theoretical understandings. Single colony of *DH5α E. coli* cells was inoculated in 10mL Luria Broth (Himedia, India) and incubated at 37$^o$C and 250rpm overnight. Cells were harvested at an OD$_{600}$ of 0.6 (corresponding to $10^8$cfu/ml) thereafter by centrifugation at 5000rpm for 10min and washed once with PBS (phosphate-buffered saline, pH~7.4). Cells were further diluted by 100 times for experimentations. Dimension of the channels is larger than the bacterial dimension by at least one order of magnitude, so that bacterium can have free access within the micro-conduit. Photolithography and subsequent soft lithography were performed to fabricate the micro-channel and thereafter bonded using oxygen plasma. *E. coli*, *DH5α* (concentration: $10^6$cfu/ml) strain suspended in phosphate buffer saline (PBS, pH~7.4) was used in the experiment. In two arms of the 'Y-microchannel', bacterial suspension and dextrose solution (3mM) respectively was transported using syringe pump. A flow rate of 1$\mu$l/min was maintained to fill the micro-conduit. Ligand concentration change with time and space in 'Y-microchannel' is shown in FIG. S11.

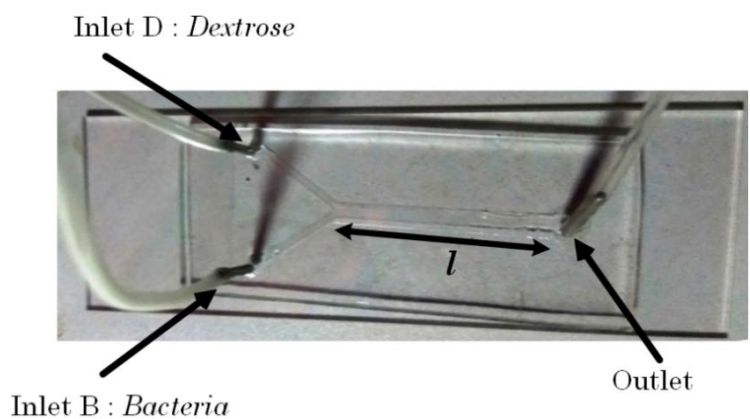

**FIG. S10:** PDMS designed 'Y-microchannel' having dimension of [length ($l$) =30 mm, width ($w$) =800 $\mu$m, and height ($h$) =25 $\mu$m].

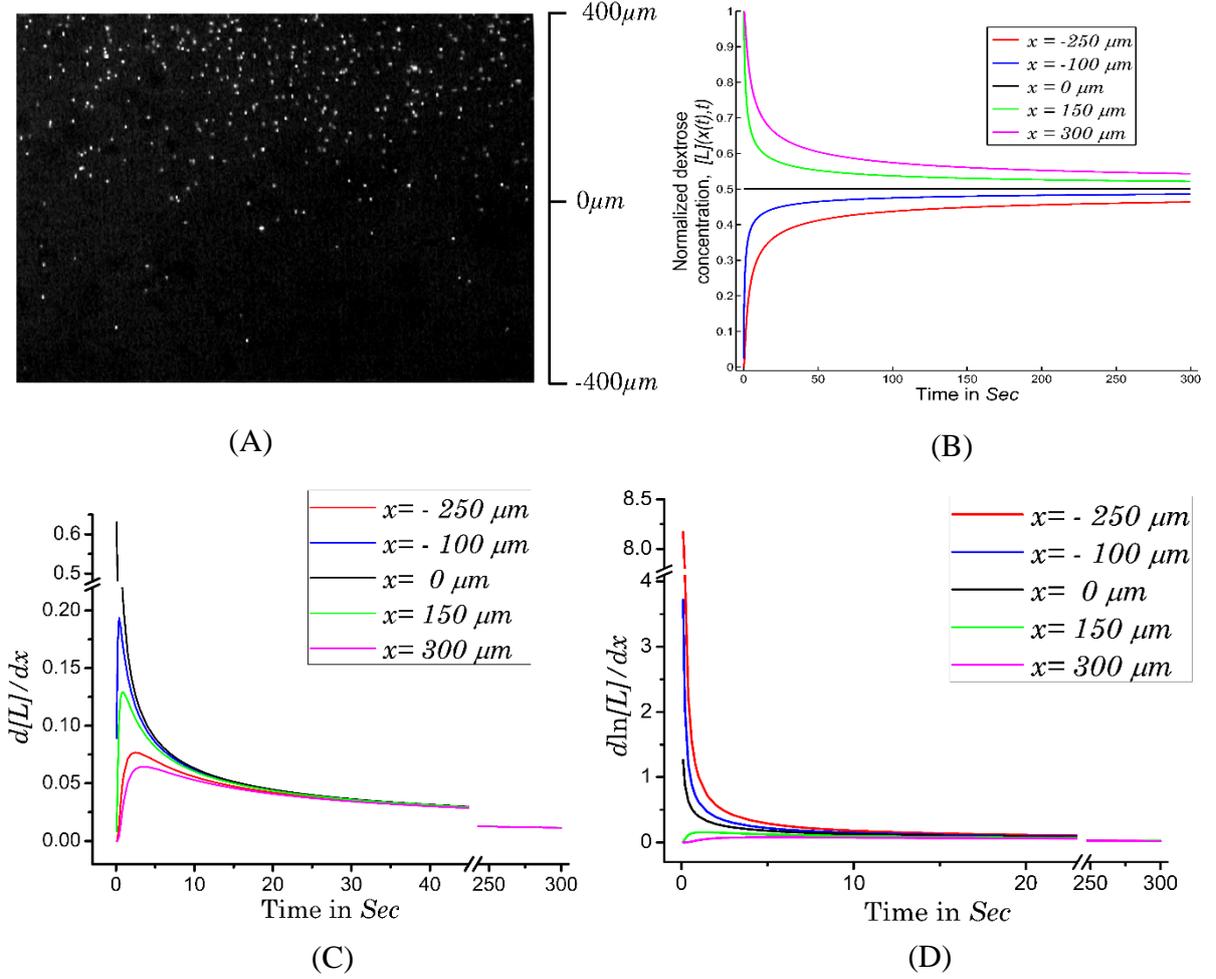

(A)

(B)

(C)

(D)

**FIG. S11:** Ligand concentration change with time and space in 'Y-microchannel'. (A) Micro-beads concentration in 'Y-microchannel': An alternate set of experiments have been performed to confirm the development of the ligand concentration profile. In these alternate experimentations, we have used a suspension of micro-beads [1 $\mu$m carboxylate-modified polystyrene fluorescent beads (purchased from Molecular Probe) suspended in PBS] instead of dextrose solution through the '*D*' reservoir and PBS solution instead of bacterial suspension in '*B*' reservoir. For chemotaxis experiment, 3 mM dextrose solution and bacteria suspension ($10^6$ cfu/ml) in PBS are transported with syringe through '*D*' and '*B*' reservoirs to microchannel and the bacteria will gain net motility towards high dextrose concentration. (B) Normalized ligand *[L](x(t),t)* concentration with time, (C) Spatial-derivative of ligand concentration $\frac{d[L](x(t),t)}{dx}$ with time, and

(D) Spatial-derivative of logarithmic ligand concentration $\frac{d}{dx}(ln[L](x(t),t))$ with time.